# The Effect of Foreign Direct Investment on Economic Growth in South Asian Countries.

S M Toufiqul Huq Sowrov

Department of Economics

University of International Business and Economics

2019

Author Note

S M Toufiqul Huq Sowrov https://orcid.org/0009-0008-6084-1393

Author is pursuing my PhD in Public Administration at NC State University (2025)

I have no known conflict of interest to disclose

For correspondence concerning this article: S M Toufiqul Huq, PhD student, Department of Public Administration, NC State University, Caldwell Hall, Raleigh, NC 27693

Sowrov07@gmail.com

shuqsow@ncsu.edu

# Abstract

This study investigates the impact of Foreign Direct Investment (FDI) on economic growth in South Asian countries, utilizing annual panel data from five SAARC member states (Bangladesh, India, Nepal, Pakistan, and Sri Lanka) over the period 1980–2017. Data sourced from the World Development Indicators and Penn World Table were analyzed using static panel models, including Ordinary Least Squares, Fixed Effects, Random Effects, and Generalized Least Squares regressions. The empirical findings reveal that FDI exhibits a consistently positive but statistically insignificant correlation with economic growth across all model specifications. In contrast, domestic investment and human capital development emerge as significant and robust positive determinants of growth. Control variables such as government consumption and inflation show expected negative, though generally insignificant, associations with growth. The results imply that for the sampled South Asian economies, enhancing domestic investment and fostering human capital are more critical for driving economic expansion than relying on FDI inflows. Consequently, policymakers should prioritize strategies that strengthen local investment climates and improve educational and skill-building institutions to boost productivity. While FDI's role remains complementary, its insignificant immediate impact suggests the need for further research into the conditional factors—such as institutional quality, financial market development, and trade policies—that might mediate its effectiveness in fostering long-term growth within the region.

# The Effect of Foreign Direct Investment (FDI) on Economic Growth in South Asian Countries

**Introduction:**

There are a lot of research works investigating causal relationship between FDI and economic growth. Some researches argue about its true effectiveness and other try to establish positive relations between flow of FDI and economic growth in low and middle income countries. Neo-classical growth and FDI-led economic development hypotheses are at the forefront of advocating and implementing pro-FDI economic policies and subsequent trade openness, whereas, mercantilist school is skeptical about FDI-led growth hypothesis and rather suggests closeness.

In this project, I tried to examine research question "The effect of FDI on the economic growth in South Asian countries". There are eight South Asian countries, Afghanistan, Bangladesh, Bhutan, India, Maldives, Pakistan and Sri Lanka. These countries are also member of a regional cooperation called SAARC which was established in 1985. I have chosen this region for my project because of my connection and the diversity it offers. The empirical study on this region exploits the available annual panel data which were collected from various sources. The data has various macroeconomic variables closely related to formation and calculation of Gross Domestic Products (GDP). The objective here is to try and test the hypothesis related to FDI led growth in South Asian countries. It was also to see the effects of FDI and domestic investment (DI) on economic growth.

**Research Question:**

The research question of this project is,

"The effect of Foreign Direct Investment (FDI) on economic growth in South Asian countries"

Therefore, null hypothesis ($H_0$) is: FDI has positive correlation with the economic growth which means inflow of FDI eventually drives up economic growth. Alternative hypothesis ($H_A$) is "no positive correlation in between."

**Literature review:**

There is always debate between pro-FDI group and "The dependency school" over the impact of foreign direct investment on economic growth. There are lot of studies and research and yet a general consensus to be reached. Sangjoon Jun (2014) tested and tried relationships between FDI and growth for the 8 SAARC countries with panel data from 1960 to 2013. His study found evidence for FDI-led growth and growth induced FDI in flow for the regional economies. Russell and Kouraklis (2015) investigated impact of FDI on growth of South Asian region by employing deductive approach and cross sectional research design and data. In their study, they found that FDI had significant impact on economic growth with the exception of Afghanistan and Pakistan. Bhavan (2011) examined the determinants and growth effect of FDI in four South Asian countries for the period of 1995-2008, with panel data. The results suggested that FDI in South Asian countries is significantly and positively associated with growth rate.

**Methodology:**

In order to study the question, static panel data model has been employed with trials of fixed effect and random effect.

1. **Data & Variables:** Panel data have been obtained from two major sources- World Development Indicators of the World Bank, for most of the data, and Penn World Table for Human Capital Index. Data range is from 1980 to 2017. Due to data insufficiency, I could not include 3 out of 8 regional countries' data. Those countries are Afghanistan, Bhutan and Maldives.

2. **Model:**

   a. <u>Baseline model:</u>

   $$Y_{it} = B_0 + B_1 FDI_{it} + B_2 DI_{IT} + B_3 GC_{IT} + B_4 inflation_{it} + B_5 HCI_{it} + B_6 lnLabor_{it} + B_7 TraOp_{it} + u_{it}$$

   b. <u>The Augmented Model:</u>

   $$Y_{it} = B_0 + B_1 FDI_{it} + B_2 DI_{IT} + B_3 GC_{IT} + B_4 inflation_{it} + B_5 HCI_{it} + B_6 lnLabor_{it} + B_7 TraOp_{it} + \gamma_1 FDIDI_{it} + \gamma_2 FDITraOp_{it} + u_{it}$$

$Y_{it}$ = percentage value of GDP for country $I$ and time $t$.
$B_o$ = Constant
$B (1-4)$ = Coefficients
FDI = Foreign Direct Investment for country $I$ and time $t$.
GC = Government Consumption for country $I$ and time $t$.
DI = Domestic Investment for country $I$ and time $t$.
HCI = Human Capital Index for country $I$ and time $t$.
Inflation = for country $I$ and time $t$.
lnLabor = Log value of total labor for country $I$ and time $t$.
TraOp = Trade Openness for country $I$ and time $t$.
U = Error term.

**<u>Regression results analysis:</u>**

To have a glimpse on overall datasets and variables nature, firstly I have presented the descriptive statistics explaining the mean, median, standard deviation, minimum and maximum values.

| Variable | Observations | Mean | Std. Dev. | Min | Max |
| GDP | 190 | 3.21578 | 2.27625 | -5.2142 | 9.00387 |
| FDI | 186 | 0.72593 | 0.72575 | -0.0984 | 3.66832 |
| DI | 190 | 22.1461 | 5.20228 | 12.5206 | 35.8129 |
| GC | 190 | 9.29971 | 2.78467 | 4.03063 | 17.6111 |
| Inflation | 190 | 8.53107 | 4.48352 | 0.15552 | 24.8912 |
| HCI | 190 | 1.79143 | 0.5041 | 1.0887 | 2.89965 |
| lnLabor | 140 | 17.4877 | 1.3939 | 15.764 | 20.0406 |
| TraOp | 190 | 40.7648 | 17.9573 | 12.2193 | 88.6364 |

**Correlation Coefficient:**

Now, I present the correlation coefficients of the variables used in this project. We see most of the variables have overall good values with only few exceptions i.e. GDP-DI and HCI-TraOP.

| | GDP | FDI | DI | GC | Inflation | HCI | lnLabor | TraOp |
|---|---|---|---|---|---|---|---|---|
| GDP | 1 | | | | | | | |
| FDI | 0.261 | 1 | | | | | | |
| DI | **0.5829** | 0.3375 | 1 | | | | | |
| GC | -0.0194 | 0.231 | -0.0398 | 1 | | | | |
| Inflation | -0.0854 | 0.1237 | -0.1668 | 0.2511 | 1 | | | |
| HCI | 0.3842 | 0.4349 | 0.372 | 0.2252 | 0.0992 | 1 | | |
| lnLabor | 0.1241 | 0.1879 | 0.3172 | 0.0197 | -0.1918 | -0.366 | 1 | |
| TraOp | 0.2561 | 0.3027 | 0.3071 | 0.2823 | 0.1639 | **0.6961** | **-0.5777** | 1 |

**Regression Results:**

| Dependent Variable: GDP (Economic growth) | | | | | |
|---|---|---|---|---|---|
| VARIABLES | (1) | (2) | (3) | (4) | (5) |
| | OLS | FE | RE | RE-Robust | GLS |
| | | | | | |
| FDI | 0.642 | 0.563 | 0.642 | 0.642 | 0.642 |
| | (0.894) | (0.989) | (0.894) | (1.165) | (0.860) |
| DI | 0.215*** | 0.140* | 0.215*** | 0.215*** | 0.215*** |

|              | (0.0606) | (0.0782) | (0.0606) | (0.0405) | (0.0584) |
|--------------|----------|----------|----------|----------|----------|
| GC           | -0.0460  | 0.00328  | -0.0460  | -0.0460  | -0.0460  |
|              | (0.0654) | (0.101)  | (0.0654) | (0.0373) | (0.0629) |
| Inflation    | -0.00254 | 0.0125   | -0.00254 | -0.00254 | -0.00254 |
|              | (0.0364) | (0.0366) | (0.0364) | (0.0559) | (0.0350) |
| HCI          | 1.062**  | 2.607    | 1.062**  | 1.062*   | 1.062**  |
|              | (0.504)  | (2.485)  | (0.504)  | (0.583)  | (0.486)  |
| lnLabor      | 0.122    | 0.778    | 0.122    | 0.122    | 0.122    |
|              | (0.214)  | (2.030)  | (0.214)  | (0.177)  | (0.206)  |
| TraOp        | 0.00225  | -0.00713 | 0.00225  | 0.00225  | 0.00225  |
|              | (0.0219) | (0.0238) | (0.0219) | (0.0322) | (0.0211) |
| FDIDI        | -0.0258  | -0.0298  | -0.0258  | -0.0258  | -0.0258  |
|              | (0.0390) | (0.0393) | (0.0390) | (0.0332) | (0.0375) |
| FDITraOp     | -0.00148 | 0.00508  | -0.00148 | -0.00148 | -0.00148 |
|              | (0.0158) | (0.0182) | (0.0158) | (0.0242) | (0.0152) |
| Constant     | -5.180   | -18.17   | -5.180   | -5.180   | -5.180   |
|              | (3.975)  | (31.94)  | (3.975)  | (4.133)  | (3.826)  |
|              |          |          |          |          |          |
| Observations | 136      | 136      | 136      | 136      | 136      |
| R-squared (within)  | 0.3801 | 0.1955 | 0.1637 | 0.1637 |  |
| R-squared (between) |        | 0.6415 | 0.9728 | 0.9728 |  |
| Number of Country   | 5      | 5      | 5      | 5      | 5 |

Standard errors in parentheses
*** $p<0.01$, ** $p<0.05$, * $p<0.1$

To identify the effects of FDI on economic growth of SAARC countries, I have run four regression methods, i.e. Ordinary Least Squares, Fixed Effects, Random Effects and Generalized Least Squares (GLS). In the first regression, mentioned above, we see our variables of interest (FDI) showed insignificant effect on economic growth but it showed the positive correlation between FDI and growth. Throughout the OLS domestic investment (DI) was significant at 1% level and Human Capital Index (HCI) was significant at 5% level. These two variables showed positive correlations with economic growth.

The second regression, FE, showed that only DI was significant at 10% level while all the other variables were insignificant. Our interest variable, FDI, showed positive correlation with growth though it's insignificant.

Third regression, RE, showed that our variable of interest (FDI) showed insignificant effect on economic growth but it showed the positive correlation between FDI and growth. Throughout the RE model, domestic investment (DI) was significant at 1% level and Human Capital Index (HCI) was significant at 5% level. These two variables showed positive correlations with economic growth. As inflation measures the level of economic stability, it shows negative correlation with economic growths of South Asian countries by theory. Herewith, we saw inflation has negative correlation with growth but not significant.

Fourth regression shows the robust estimates of RE. Only change was in HCI significance level and it was significant at 10% level. No other changes found from the robust estimates of RE. It was done in order to remove heteroskedasticity.

The Fifth regression line, Generalized Least Squares (GLS), removed the problem of heteroskedasticity and autocorrelation. It showed that domestic investment (DI) was significant at 1% level and Human Capital Index (HCI) was significant at 5% level. These two variables showed positive correlations with economic growth. As inflation measures the level of economic stability, it shows negative correlation with economic growths of South Asian countries by theory. Herewith, we saw inflation has negative correlation with growth but not significant.

Our interest variable, FDI, showed positive correlations in all measures but was not significant in any of those regressions.

**Conclusion and Recommendation:**

This project explores the correlations between FDI and economic growths in South Asian countries by using static panel data model. This empirical study exploits annual panel data on five South Asian countries' macroeconomic variables. The annual panel data are extracted from the World Development Indicators (WDI) 2018 of the World Bank, and the sample period runs from 1980 to 2017. The paper, which explores the causal link between foreign direct investment and economic growth, are as follows.

All five regressions showed positive but insignificant relationship between FDI and economic growth of South Asian countries but it is also noticeable that DI became significant in all cases ranging from 1% to 10% level, whereas GC and Inflation were negative in almost all regressions.

The empirical findings above have some interesting policy implications for South Asian countries. An effective investment policy for encouraging more domestic investments in order to achieve more economic growth and at the same time, countries from this region should formulate policies supporting better human resources generation. Those will increase total factor of productivity (TFP) and, thus, boosting economic growth. Our interest variable, FDI, has insignificant but positive impact on economic growth for this region. Further researches can be done in this area to study broad impact of FDI to economic growth. Due to data insufficiency, thorough studies could not be done on all the regional 8 (eight) countries. But I have tried best to produce accurate results from the available data with various regressions. The results suggest that although FDI showed insignificant in all respects, we cannot ignore its importance to augment economic growth. So, South Asian countries should have comprehensive policies to utilize its human capital and to control inflation over the time to boost economic growth.

# **Annexures**

## OLS result

`. reg GDP FDI DI GC Inflation HCI lnLabor TraOp FDIDI FDITraOp`

| Source | SS | df | MS | | |
|---|---|---|---|---|---|
| Model | 233.404511 | 9 | 25.9338346 | | |
| Residual | 380.63402 | 126 | 3.02090492 | | |
| Total | 614.038531 | 135 | 4.54843357 | | |

Number of obs = 136
F( 9, 126) = 8.58
Prob > F = 0.0000
R-squared = 0.3801
Adj R-squared = 0.3358
Root MSE = 1.7381

| GDP | Coef. | Std. Err. | t | P>|t| | [95% Conf. Interval] | |
|---|---|---|---|---|---|---|
| FDI | .6419327 | .8936627 | 0.72 | 0.474 | -1.1266 | 2.410465 |
| DI | .2153367 | .0606313 | 3.55 | 0.001 | .0953492 | .3353242 |
| GC | -.0459845 | .0653916 | -0.70 | 0.483 | -.1753926 | .0834236 |
| Inflation | -.0025358 | .0363523 | -0.07 | 0.944 | -.074476 | .0694044 |
| HCI | 1.062422 | .5044806 | 2.11 | 0.037 | .0640698 | 2.060774 |
| lnLabor | .1222667 | .2142846 | 0.57 | 0.569 | -.3017962 | .5463295 |
| TraOp | .0022497 | .0219241 | 0.10 | 0.918 | -.0411374 | .0456369 |
| FDIDI | -.0258345 | .0389816 | -0.66 | 0.509 | -.1029779 | .0513089 |
| FDITraOp | -.0014752 | .0157851 | -0.09 | 0.926 | -.0327135 | .029763 |
| _cons | -5.180274 | 3.975184 | -1.30 | 0.195 | -13.04705 | 2.686499 |

## FE result

`. xtreg GDP FDI DI GC Inflation HCI lnLabor TraOp FDIDI FDITraOp, fe`

Fixed-effects (within) regression  Number of obs = 136
Group variable: Country  Number of groups = 5

R-sq: within = 0.1955  Obs per group: min = 24
between = 0.6415  avg = 27.2
overall = 0.2950  max = 28

F(9,122) = 3.29
corr(u_i, Xb) = -0.6577  Prob > F = 0.0013

| GDP | Coef. | Std. Err. | t | P>|t| | [95% Conf. Interval] | |
|---|---|---|---|---|---|---|
| FDI | .5629006 | .9893837 | 0.57 | 0.570 | -1.395683 | 2.521484 |
| DI | .1396842 | .0781783 | 1.79 | 0.076 | -.0150776 | .294446 |
| GC | .003281 | .1010709 | 0.03 | 0.974 | -.1967989 | .2033609 |
| Inflation | .0124852 | .0366011 | 0.34 | 0.734 | -.0599704 | .0849407 |
| HCI | 2.607476 | 2.485201 | 1.05 | 0.296 | -2.312227 | 7.52718 |
| lnLabor | .7775892 | 2.030353 | 0.38 | 0.702 | -3.241697 | 4.796875 |
| TraOp | -.0071292 | .0237695 | -0.30 | 0.765 | -.0541834 | .039925 |
| FDIDI | -.0297746 | .0392947 | -0.76 | 0.450 | -.1075625 | .0480132 |
| FDITraOp | .0050826 | .0182143 | 0.28 | 0.781 | -.0309745 | .0411396 |
| _cons | -18.17434 | 31.93611 | -0.57 | 0.570 | -81.39505 | 45.04637 |
| sigma_u | 1.1034588 | | | | | |
| sigma_e | 1.7231803 | | | | | |
| rho | .29081192 | (fraction of variance due to u_i) | | | | |

F test that all u_i=0: F(4, 122) = 1.55  Prob > F = 0.1929

# RE result

```
. xtreg GDP FDI DI GC Inflation HCI lnLabor TraOp FDIDI FDITraOp, re

Random-effects GLS regression                  Number of obs      =       136
Group variable: Country                        Number of groups   =         5

R-sq:  within  = 0.1637                        Obs per group: min =        24
       between = 0.9728                                       avg =      27.2
       overall = 0.3801                                       max =        28

                                               Wald chi2(9)       =     77.26
corr(u_i, X)   = 0 (assumed)                   Prob > chi2        =    0.0000
```

| GDP | Coef. | Std. Err. | z | P>\|z\| | [95% Conf. Interval] | |
|---|---|---|---|---|---|---|
| FDI | .6419327 | .8936627 | 0.72 | 0.473 | -1.109614 | 2.39348 |
| DI | .2153367 | .0606313 | 3.55 | 0.000 | .0965016 | .3341718 |
| GC | -.0459845 | .0653916 | -0.70 | 0.482 | -.1741497 | .0821807 |
| Inflation | -.0025358 | .0363523 | -0.07 | 0.944 | -.0737851 | .0687135 |
| HCI | 1.062422 | .5044806 | 2.11 | 0.035 | .0736582 | 2.051186 |
| lnLabor | .1222667 | .2142846 | 0.57 | 0.568 | -.2977234 | .5422567 |
| TraOp | .0022497 | .0219241 | 0.10 | 0.918 | -.0407207 | .0452202 |
| FDIDI | -.0258345 | .0389816 | -0.66 | 0.507 | -.102237 | .050568 |
| FDITraOp | -.0014752 | .0157851 | -0.09 | 0.926 | -.0324135 | .029463 |
| _cons | -5.180274 | 3.975184 | -1.30 | 0.193 | -12.97149 | 2.610944 |
| sigma_u | 0 | | | | | |
| sigma_e | 1.7231803 | | | | | |
| rho | 0 | (fraction of variance due to u_i) | | | | |

## RE Robust

```
. xtreg GDP FDI DI GC Inflation HCI lnLabor TraOp FDIDI FDITraOp, re r

Random-effects GLS regression                   Number of obs      =        136
Group variable: Country                         Number of groups   =          5

R-sq:  within  = 0.1637                         Obs per group: min =         24
       between = 0.9728                                        avg =       27.2
       overall = 0.3801                                        max =         28

                                                Wald chi2(4)       =          .
corr(u_i, X)   = 0 (assumed)                    Prob > chi2        =          .

                            (Std. Err. adjusted for 5 clusters in Country)
```

|          |             | Robust     |       |       |                       |
|----------|-------------|------------|-------|-------|-----------------------|
| GDP      | Coef.       | Std. Err.  | z     | P>\|z\| | [95% Conf. Interval] |
| FDI      | .6419327    | 1.164781   | 0.55  | 0.582 | -1.640996   2.924862 |
| DI       | .2153367    | .0404501   | 5.32  | 0.000 | .136056    .2946174  |
| GC       | -.0459845   | .0373061   | -1.23 | 0.218 | -.1191032   .0271342 |
| Inflation| -.0025358   | .0558766   | -0.05 | 0.964 | -.1120519   .1069803 |
| HCI      | 1.062422    | .5832076   | 1.82  | 0.069 | -.0806439   2.205488 |
| lnLabor  | .1222667    | .1766881   | 0.69  | 0.489 | -.2240356   .4685689 |
| TraOp    | .0022497    | .0321507   | 0.07  | 0.944 | -.0607645   .065264  |
| FDIDI    | -.0258345   | .0332088   | -0.78 | 0.437 | -.0909226   .0392536 |
| FDITraOp | -.0014752   | .0242296   | -0.06 | 0.951 | -.0489643   .0460138 |
| _cons    | -5.180274   | 4.133207   | -1.25 | 0.210 | -13.28121   2.920662 |
| sigma_u  | 0           |            |       |       |                      |
| sigma_e  | 1.7231803   |            |       |       |                      |
| rho      | 0           | (fraction of variance due to u_i) |       |       |          |

## GLS

```
. xtgls GDP FDI DI GC Inflation HCI lnLabor TraOp FDIDI FDITraOp

Cross-sectional time-series FGLS regression

Coefficients:  generalized least squares
Panels:        homoskedastic
Correlation:   no autocorrelation

Estimated covariances      =         1          Number of obs      =         136
Estimated autocorrelations =         0          Number of groups   =           5
Estimated coefficients     =        10          Obs per group: min =          24
                                                               avg =        27.2
                                                               max =          28
                                                Wald chi2(9)       =       83.40
Log likelihood             = -262.9601          Prob > chi2        =      0.0000
```

| GDP | Coef. | Std. Err. | z | P>\|z\| | [95% Conf. Interval] | |
|---|---|---|---|---|---|---|
| FDI | .6419327 | .8601803 | 0.75 | 0.456 | -1.04399 | 2.327855 |
| DI | .2153367 | .0583596 | 3.69 | 0.000 | .1009539 | .3297195 |
| GC | -.0459845 | .0629416 | -0.73 | 0.465 | -.1693478 | .0773788 |
| Inflation | -.0025358 | .0349904 | -0.07 | 0.942 | -.0711156 | .066044 |
| HCI | 1.062422 | .4855795 | 2.19 | 0.029 | .1107038 | 2.01414 |
| lnLabor | .1222667 | .2062561 | 0.59 | 0.553 | -.2819878 | .5265211 |
| TraOp | .0022497 | .0211027 | 0.11 | 0.915 | -.0391107 | .0436102 |
| FDIDI | -.0258345 | .0375211 | -0.69 | 0.491 | -.0993745 | .0477054 |
| FDITraOp | -.0014752 | .0151937 | -0.10 | 0.923 | -.0312543 | .0283039 |
| _cons | -5.180274 | 3.826248 | -1.35 | 0.176 | -12.67958 | 2.319034 |